\documentclass[11pt,a4paper]{article}
\usepackage[T1]{fontenc}
\usepackage[utf8]{inputenc}
\usepackage{amsmath}
\usepackage{amsfonts}
\usepackage{amssymb}
\usepackage{graphicx}
\usepackage{subfig}
\usepackage{color}
\usepackage{titlesec}

\title{\vspace*{-1em}\Large\bf Spontaneous currents in a bosonic ring}
\author{\normalsize Damian Makieła and Maciej M. Ma\'{s}ka\footnote{E-mail: 
\tt{maciej.maska@phys.us.edu.pl}}\\
\normalsize Department of Theoretical Physics, Institute of Physics, University of Silesia\\ 
\normalsize ul. Uniwersytecka 4, 40-007 Katowice, Poland}
\date{}
\makeatother
\titlespacing\section{0pt}{10pt plus 3pt minus 3pt}{5pt plus 2pt minus 2pt}
\titleformat{\title}
  {\normalfont\fontsize{12}{15}\bfseries}{\thesection}{1em}{}
\titleformat{\section}
  {\normalfont\fontsize{12}{15}\bfseries}{\thesection}{1em}{}
\renewcommand{\thesection}{\arabic{section}.}

\begin{document}
\maketitle
\begin{abstract}
  Nonequilibrium dynamics of noninteracting bosons in a one-dimensional ring-shaped lattice
  is studied by means of the Kinetic Monte Carlo method. The system is approximated
  by the classical XY model (the kinetic term is neglected) and then the 
  simulations are performed for the planar classical spins. 
  We study the dynamics that follows a finite-time quench to zero temperature. If the quench
  is slow enough the system can equilibrate and finally reaches the ground state with uniform
  spin alignment. However, we show that if the quench is faster than the relaxation
  rate, the system can get locked in a current-carrying metastable state characterized by a
  nonzero winding number. We analyze how the zero-temperature state depends on the quench 
  rate.
\end{abstract}	

PACS 03.75.Lm, 03.75.Kk, 64.60.Ht

\section{Introduction}

It has been predicted many years ago that small metallic rings may have a symmetry-breaking
ground state with spontaneous orbital currents, i.e., currents occurring in the absence of
applied magnetic field or power sources \cite{Wohl}. Similar behaviour was also expected as
a result of the parity effect in superconducting nanorings \cite{parity} or chiral tunneling
in carbon nanotubes \cite{nanotube}. Rotating ground state was predicted in a one-dimensional
spin-polarized gas composed of an even number of fermionic atoms interacting via attractive
$p$-wave interactions and confined to a mesoscopic ring \cite{Girardeau}. In 1977 Bulaevskii
{\em et al.} proposed that current can flow in the ground state of a superconducting ring
with $\pi$-junction \cite{bulaevskii}. Such currents have recently been observed in
superconducting networks \cite{Frolov}. Realization of this idea in superfluid fermionic cold
gases in a ring-shaped trap has been proposed \cite{pi}. 

In this paper we demonstrate that a similar phenomenon can be expected also in a system of
free non-interacting bosons in a one-dimensional ring. The difference with respect to the
mentioned above examples is that in our case the current flows in a metastable state, not in
the ground state. We show, however, that the system can be easily trapped in such a state by
a sufficiently fast temperature quench that leads to a non-uniform symmetry breaking.

Non-uniform symmetry breaking was studied by Kibble \cite{kibb1}, who has shown that the
cooling down of the early universe resulted in independent symmetry breakings in distant
regions. This, in turn, led to the formation of topological defects such as point-like
monopoles, linear cosmic strings, or planar domain walls. Structures analogous to cosmic
strings exist in many condensed matter systems \cite{zurek}. Some examples
of these defects are magnetic flux tubes in superconductors, vortices in superfluids, and
certain defects in liquid crystals. The theory containing a complex field $\phi$ constitutes
the simplest example which has a string analogues. The mean value of $\phi$ above the
critical temperature is equal to zero. Below this temperature, $\phi$ acquires a non-zero
value that causes symmetry breaking as a consequence of a phase transition.
One of the examples of such systems is superfluid helium-4. At low temperature a significant
part of atoms of superfluid helium-4 occupies a single quantum state. Then, $\phi$ is
proportional to the wave function. A vortex is a topological defect around which the phase
of $\phi$ changes by $2\pi$. There occurs quantized circulation around the vortex because
the superfluid velocity is proportional to the phase gradient. In the case of
superconductors, the quantized quantity is the magnetic flux carried by the vortex.

One of the first experiments in which a rapid transition led to the formation of linear
defects was done on nematic liquid crystals formed of rod-shaped molecules \cite{kibb3}.
Another example is the experiments performed on superfluid helium-3 at the
temperature around 2 mK. Helium-3 reacts with slow neutrons as
${}^1_0n + {}^3_2He \rightarrow {}^1_1p + {}^3_1H$, releasing 764 keV of energy.
In experiments where this reaction was used the neutron absorption heated up a small area
of superfluid helium-3 above the critical temperature \cite{kibb4,kibb5}. Fast cooling of
this area below the transition temperature led to the formation of vortices. In an
experiment performed at CNRS Center for Research on Ultra-Low Temperatures in Grenoble,
France, \cite{kibb4} the total energy released after neutron absorption was investigated.
It was found that observed energy deficit can be explained by the formation of vortices.

Several experiments were carried out also on superconductors. Some of them involved thin
films of high temperature superconductors, first heated above the critical temperature,
and then cooled back through the phase transition. The generated defects, which are tubes
of quantized magnetic flux passing through the film in one direction or the other, are
called fluxons and antifluxons \cite{kibb8}. Researchers measured the net defect number
$\Delta N = N_{+}-N_{-}$ where $N_{+}$ and $N_{-}$ are the numbers of fluxons
and antifluxons, respectively. John Kirtley and others performed experiments on thin-film 
superconducting rings \cite{kibb11}. 

Interesting experiments were performed also with Josephson junctions. Monaco and 
collaborators used the annular Josephson junctions \cite{kibb12}. The rapid cooling 
generated a flux in the system. Then, the dependence of the probability of catching 
fluxon on the quench rate was measured.
Cold atomic gases are another type of system in which there are vortices. Their advantage
is high purity and theoretical simplicity \cite{kibb13}.

The experiments mentioned above are evidences of the fact that a rapid phase transition
leads to the creation of topological defects. Below we demonstrate how such a process 
can be quantitatively studied in a simple model of noninteracting itinerant bosons in a 
one-dimensional lattice with periodic boundary conditions.

\section{Model Hamiltonian}

We start with a Hamiltonian that describes the kinetic energy of noninteracting bosons, 
given by

\begin{equation}
H=t\sum_{\langle i,j\rangle}b_{i}^{\dagger}b_{j},
\label{bosons}
\end{equation}
where $b_{i}\ (b_i^\dagger)$ is an operator that annihilates (creates) a boson in site $i$,  
$t$ is the hopping amplitude, $\left\langle i,j\right\rangle$ denotes summation
over nearest neighbouring sites. Using the approximation
\begin{equation}
b_{i}\longrightarrow\sqrt{n_{i}}e^{i\theta_{i}},\quad n_i=n=\mathrm{const},
\label{approx}
\end{equation}
Hamiltonian (\ref{bosons}) can be rewritten as the XY Hamiltonian

\begin{equation}
H=J\sum_{\langle i,j\rangle}\cos(\theta_{i}-\theta_{j}),
\label{XY}
\end{equation}
where $J=2tn$. Approximation (\ref{approx}), where bosonic operators are replaced by 
$C$-numbers, neglects quantum fluctuations and generally is not valid.\footnote{Without
approximations Hamiltonian (\ref{bosons}) can be mapped onto the two-dimensional XY 
model \cite{caz}.} However, with increasing $n$ the fluctuations becomes small. 
The Hamiltonian (\ref{XY}) is particularly justified if there are Bose-Einstein
condensates in all lattice sites. Such systems have been realized experimentally in 
cold atoms \cite{ringBEC}. 

\begin{figure}[htb]
\centerline{\includegraphics[width=0.8\linewidth]{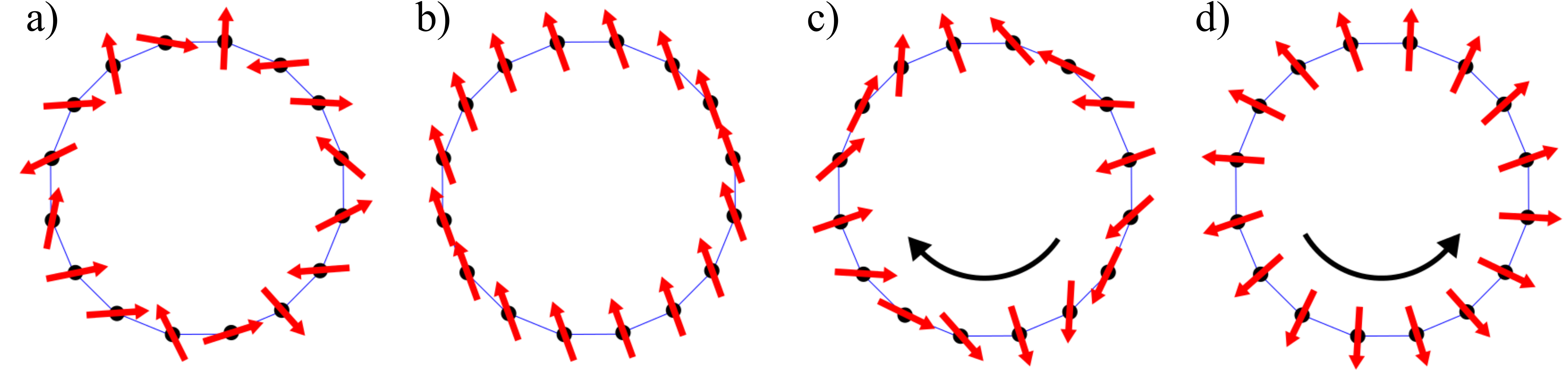}}
\caption{Spin configurations in the one-dimensional ring-shaped lattice at high
  temperature (a) and at $T=0$ (b, c, and d). Panel b) shows configuration with $W_N=0$,
whereas panels c) and d) configurations with $W_N=+1$ and $W_N=-1$, respectively.}
\label{fig0}
\end{figure}

Complex variables $\theta_i$ in Eq. (\ref{XY}) can can be interpreted as directions 
of planar spins with their dynamics governed be interaction 
$J\cos(\theta_{i}-\theta_{j})$.
At high temperature they are are random. However, with decreasing temperature the
interaction energy try to align the spins in the same direction. The one-dimensional 
XY model does not have a long range order at any finite temperature, so we can expect 
the fully polarized state only at $T=0$ (see Fig. \ref{fig0}a and \ref{fig0}b).

The interaction energy is minimal if a given spin points in the same 
direction as its neighbours. If the system is approaching the symmetry-breaking
phase slowly, the spins can evolve almost adiabatically eventually reaching the
global minimum of energy, i.e, the configuration presented in Fig. \ref{fig0}a. 
However, if the temperature quench is sufficiently fast, the system can end up in a
local minimum of energy where spins are (almost) parallel only locally. Since the 
phase $\theta_i$ must be a single-valued function of position $i$, its total change 
along a closed path must be equal to $2\pi W_N$, where $W_N$ is an integer. Then,
all spin configurations can be classified by the value of $W_N$ and only the one
with $W_N=0$ corresponds to the global minimum of energy. Since $W_N$'s are 
integer, configurations with different $W_N$ cannot be continuously transformed
one into another. $W_N$ is named a {\em winding number} and can formally be 
defined as

\begin{equation}
W_{N}=\frac{1}{2\pi}\sum_{\langle i,j\rangle} (\theta_{i}-\theta_{j}).
\end{equation}

Finite value of the winding number requires a non-zero gradient of the boson wave 
function and indicates the existence of a current flowing along the ring \cite{mon}.
Examples of such configurations are presented in Fig. \ref{fig0}a and \ref{fig0}b.

The question we want to address in this paper is how the value of $W_N$ depends
on the cooling rate. In the case of the second order phase transition the answer 
is suggested by the famous Kibble-\.Zurek hypothesis \cite{kibb1,zurek}.  

\section{Simulations of the spin dynamics}

Here, we propose to perform a computer experiment to ``measure'' $W_N$ for 
different cooling rates. The most natural approach would be to use the Monte Carlo 
(MC) method to simulate the the behaviour of the system during the temperature 
quench. The problem, however, is that in the standard MC approaches there is no 
relation between the number of MC steps and the real time. This relation is 
particularly disturbed close to a phase transition where the {\em critical slowing
down} occurs. In order to overcome these difficulties we propose to use the {\em Kinetic
Monte Carlo} (KMC) method. It is a method commonly used to study time dependence 
of processes occurring in nature. If these processes have known
transition rates between different states, KMC allows one to determine the 
relation between the number of steps in the algorithm and the real
time of the process. In the case of the XY model the rate is defined by the inverse
of the coupling $J$.

\begin{figure}[htb]
\centerline{\includegraphics[width=0.32\linewidth]{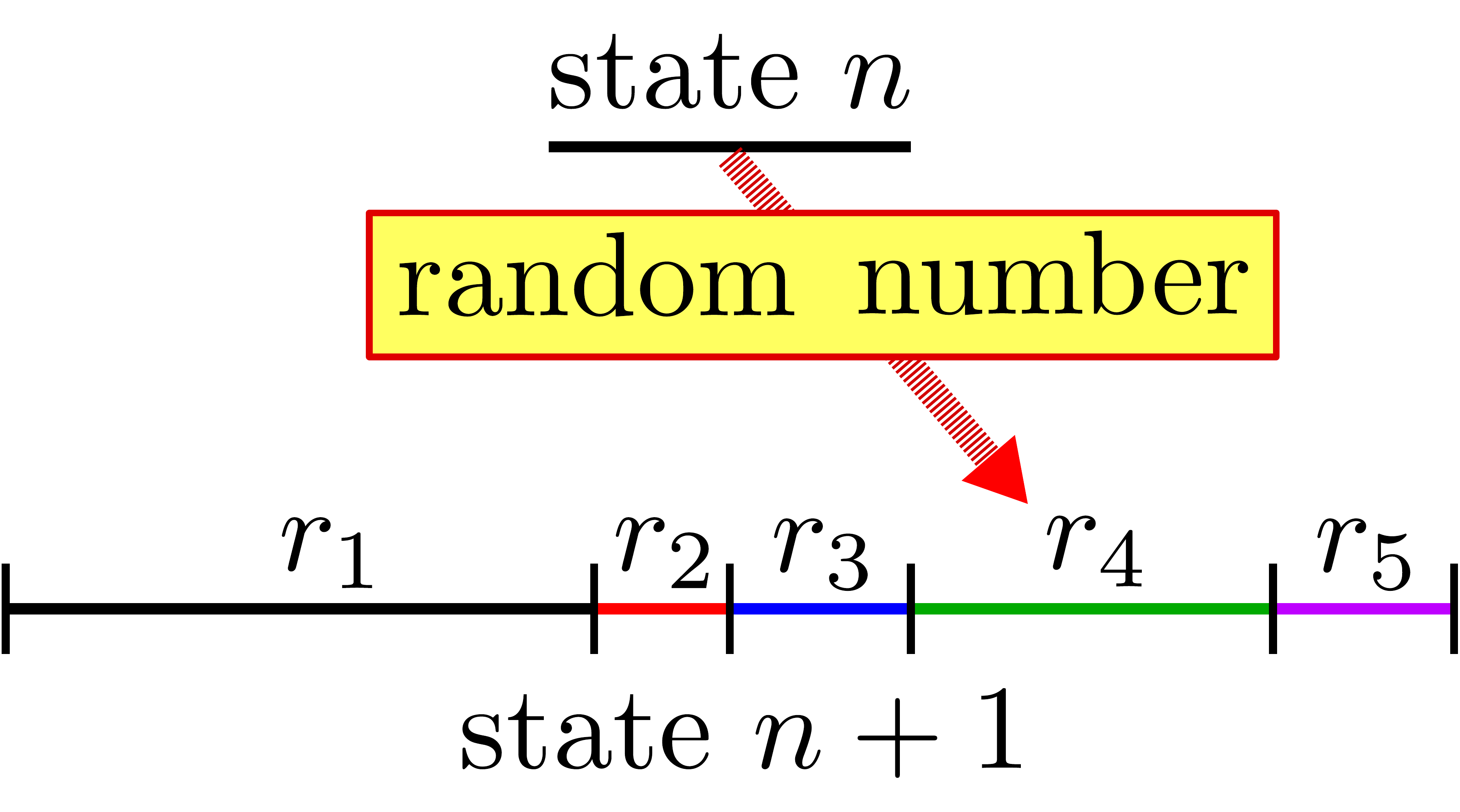}}
\caption{Illustration of the Kinetic Monte Carlo method. The probability that in a given 
MC step the system will evolve to state $i$ is proportional to $r_i$. See the text 
for detailed explanation.}
\label{fig2}
\end{figure}

The computer ``experiments'' are performed as follows: 
For a given number of lattice sites $N$ we chose an initial random high-temperature 
configuration of the spins. Then, the temperature decreases linearly with time to zero. 
The quench rate is defined by $\tau_Q$:
\begin{equation}
  T(t)=\left\{\begin{array}{ll}
-T_{0}\displaystyle\frac{t}{\tau_{Q}} & \mathrm{for}\ t \in[-\tau_{Q},0),\\[0.3em]
0 & \mathrm{for}\ t \in[0,t_R),
\end{array}\right.
  \label{cooling_rate}
\end{equation}
where time $t_R$ is sufficiently long to allow the system to relax to its local 
energy minimum after the temperature quench. The evolution of the system is 
defined by the following algorithm sketched in Fig. \ref{fig2}.

At each step of simulation we must consider {\em all} possible ways of evolution to 
the next state. The corresponding transition probabilities $r_{i}$ are given by 
the change of energy $\Delta E$ between the states and the current 
temperature of the system, $r_i~\propto~\exp(-\Delta E_i/k_BT)$.  
Next, we compute
the cumulative function $R_{i}=\sum_{n=1}^{i}r_{n}, \quad i=1,...,M$, where $M$ is the number
of all the states that can be directly reached from the present state, and generate a random
number $q$ from the uniform distribution  $(0,R_{N}]$. We find the index $j$ satisfying
inequality $R_{j-1} < q \le R_{j}$. It means that the number $q$ is in the field  $r_{j}$
(Fig.~\ref{fig2}) and the system will go to state $j$. The main numerical difficulty
in this method
is that the number of possible transitions from a given state increases exponentially 
with the number of lattice sites. 

\begin{figure}[htp]
\centering
\includegraphics[width=\linewidth]{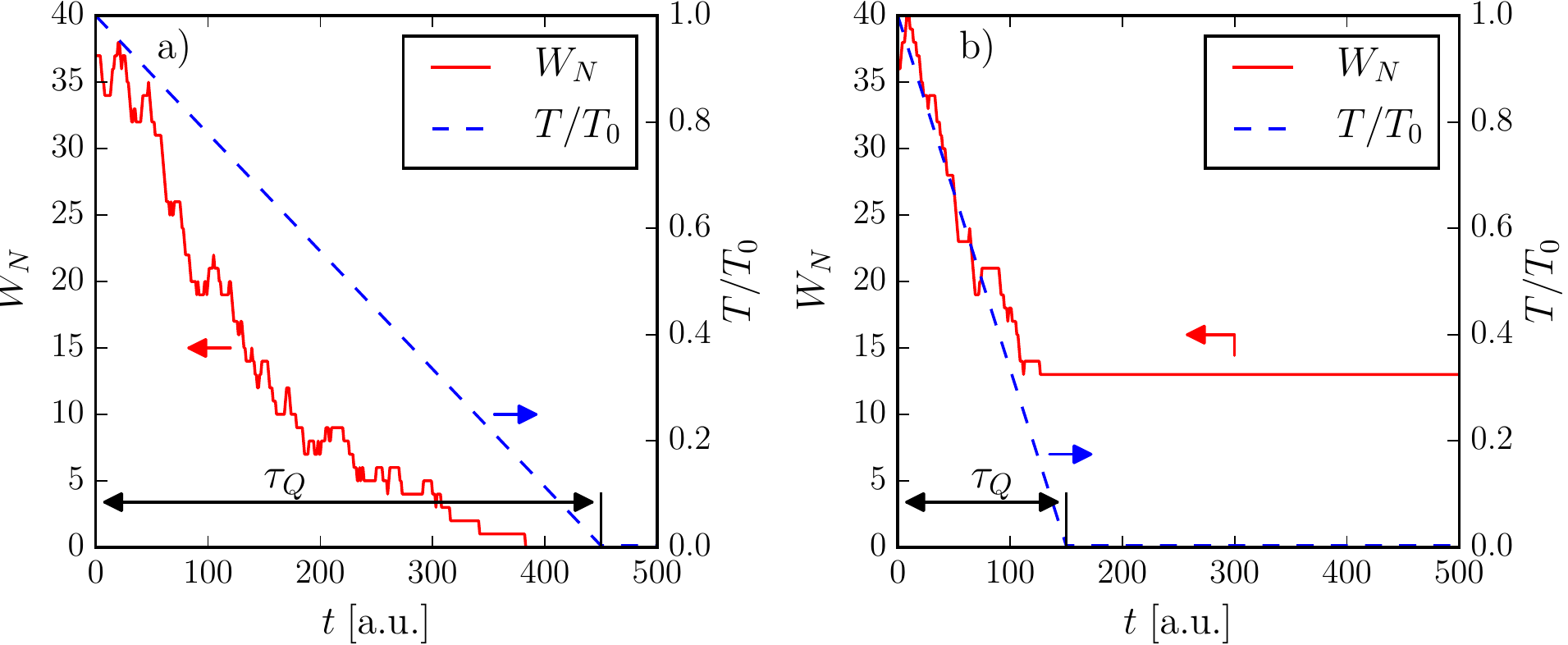}
\caption{The time dependence of the
  winding number $W_{N}$ for slow (a) and fast (b) cooling. The solid red line shows the 
time evolution of $W_N$, whereas the dashed blue line shows the temperature. The black 
horizontal arrow indicates the value of the cooling rate $\tau_Q$ 
[Eq. (\ref{cooling_rate})]. Panel b) illustrates a situation where after the quench the 
system remains in a state with non-zero winding number.}
\label{fig3}
\end{figure}

After the quench the system evolves to its final configuration, and then we ``measure''
the winding number. Fig. \ref{fig3}a and \ref{fig3}b show the evolution in the case
of slow and fast temperature quench, respectively.  

The results show that when the cooling is slow the system is able to maintain its quasi 
adiabatic evolution and eventually ends up in the state corresponding to the global 
minimum of energy with $W_N=0$. On the other hand, if the quench is faster than the 
relaxation rate, the system can get locked in a metastable current-carrying state with 
non-zero $W_N$. Detailed analysis of the results show that the dependence of the average 
value of $W_N$ on the quench rate $\tau_Q$ and on the system size $N$ is in agreement 
with predictions of the Kibble-\.Zurek hypothesis.
  
\section{Conclusions}

We have shown that the temperature quench in a bosonic ring can produce {\em spontaneous} 
currents. The most natural system where such currents could be observed is a ring-shaped
optical lattice with Bose-Einstein condensates in each lattice site \cite{ami,ringBEC}. 
It is, however, 
possible that similar effects may be realized by temperature quench also in other systems 
described by the one-dimensional XY model, e.g., granular superconducting rings or rings 
of Josephson junctions \cite{ringJJ}. 

\section*{Acknowledgment} 
This work was supported by the Polish National Science Centre (NCN) under grant
DEC-2013/11/B/ST3/00824.

\end{document}